\documentclass{aa}
\usepackage{txfonts}
\usepackage{natbib}
\bibpunct{(}{)}{;}{a}{}{,} 
\usepackage{graphicx}
\usepackage{color}
\usepackage{hyperref}
\usepackage{ulem}

\begin{document}

\title{Star formation and environment in clusters up to $z \sim 2.2$}
\author{A. Raichoor  \and S. Andreon}
\institute{INAF -- Osservatorio Astronomico di Brera, via Brera 28, 20121 Milan, Italy\\
e-mails: \url{anand.raichoor@brera.inaf.it}, \url{stefano.andreon@brera.inaf.it}}

\date{Accepted ... . Received ...}

\abstract
{The dependence of galaxy star formation activity on environment -- especially in clusters -- at high redshift is still poorly understood, as illustrated by the still limited number of $z \gtrsim 1.4$ clusters on the one hand, and by the still debated star formation-density relation at high redshift on the other hand.
}
{The $z_{phot} \sim 2.2$ JKCS\,041 cluster allows to probe such environmental dependence of star formation activity at an unprecedented combination of redshifts and environments.
Its study permits to enlarge the knowledge of high redshift clusters and to put strong leverage on observational constraints for galaxy evolution models.
}
{We analyze deep $u^*g'r'i'z'JHK_s$ images from the CFHTLS/WIRDS surveys, which cover JKCS\,041 cluster field.
We first estimate photometric redshifts based on multi-wavelength photometry.
We then lead a careful analysis to test the presence of a Butcher-Oemler effect.
We work on galaxies within $2 \times r_{200}$ and with masses $\ge 1.34 \times 10^{11} M_\odot$, and use two comparison clusters at $z = 0$ and $z = 1$ of similar mass.
We estimate the radial profiles of the fraction of blue galaxies, taking into account the star aging with decreasing redshift.
}
{After confirming the high redshift nature of JKCS\,041, we find no evidence for a Butcher-Oemler effect between $z \sim 2.2$ and $z \sim 0$ for galaxies more massive than $1.34 \times 10^{11} M_\odot$.
In the cluster center, a change greater than $\Delta  f_{blue} / \Delta z = 0.16$ between $z \sim 0$ and $z \sim 2.2$ would be easily detected.
We also find that JKCS\,041 shows a consistent and systematic increase of the fraction of star-forming galaxies with cluster-centric distance, hence with decreasing density, for both a $M \ge 1.34 \times 10^{11} M_\odot$ selected sample and a lower mass sample.
In particular, very few (less than 15\%) star-forming galaxies are found within $r_{200}/2$ among high mass ($M \ge 1.34 \times 10^{11} M_\odot$) galaxies.
}
{Our results show that the present-day star formation-density relation is already in place at $z \sim 2.2$.
}

\titlerunning{Star formation and environment in clusters up to $z \sim 2.2$}
\authorrunning{A. Raichoor \& S. Andreon}
\maketitle


\section{Introduction \label{sec:intro}}

	It is known that, in the local Universe, star formation activity is correlated with galaxy environments: galaxies with low star formation rates are preferentially found in dense environment, galaxies in cluster cores being virtually all quiescent \citep[e.g.][]{oemler74,hogg03,kauffmann04}.
However the various processes which led to this situation in the local Universe are still not fully understood \citep[e.g.][]{treu03}.
To put constraints on how this star formation-density relation, well established in the local Universe, has been built through cosmic ages, one should observe galaxies at increasing look-back times and study their star formation activity with respect to the environment at fixed stellar mass, so as to isolate the role of environment.
In this context, galaxy clusters bring decisive informations, as they are the densest environments in the Universe.

	It has been shown that the star formation-density relation holds out to $z \lesssim 0.8$ \citep[e.g.][]{patel11}.
At $z \sim 0.8$-1, while studies in low-density environments \citep[e.g.][]{elbaz07,cooper08} observe a reversal of the star formation-density relation, other studies focusing on cluster environments \citep[e.g.][]{patel09a,koyama10} find that cluster core regions are devoid of star-forming galaxies.
\citet{sobral11}, studying a very wide range of environments, nicely reconcile those observations: star formation activity increases with increasing density up to $\Sigma \sim 10$-30 Mpc$^{-2}$, and then decreases with increasing density for $\Sigma \gtrsim 30$ Mpc$^{-2}$.

	Studying a superstructure at $z \sim 1.2$, \citet{tanaka09} find that the star formation-density relation is already in place.
When going to higher redshifts ($z \sim 1.4$-1.6), the situation is less clear.
On the one hand, \citet{hayashi10} and \citet{hilton10} study the XMMXCS J2215.9-1738 cluster at $z = 1.46$ \citep[$kT \sim 4.1$ keV,][]{stanford06,hilton10}: looking at the 24 $\mu$m and [OII] emission respectively, both work observe a high star formation activity at its centre.
On the other hand, studies on the massive XMMU J2235-2557 cluster at $z = 1.39$ \citep[$kT \sim 8.6$ keV,][]{jee09,rosati09} observe that its core region does not present star formation activity \citep{lidman08,rosati09,strazzullo10,bauer11}.
\citet{chuter11} and \citet{quadri11} study a large sample of galaxies and environments in the UKIDSS Ultra-Deep Survey \citep[UDS, ][; O. Almaini, in preparation]{lawrence07} and find that the star formation-density relation holds out to $z \sim 1.5$-1.8.
\citet{quadri11} also investigate the ClG J0218.3-05101 cluster at $z = 1.62$ \citep[$kT \sim 1.7$ keV,][]{papovich10,tanaka10} and observe that its central region has an elevated fraction of quiescent objects relative to the field, in apparent disagreement with \citet{tran10}, who observe an increasing of the star formation activity along with the density.

	Though the two clusters XMMU J2235-2557 and XMMXCS J2215.9-1738 lie at a similar redshift of $z \sim 1.4$, XMMU J2235-2557 is massive and presents a well defined red sequence down to faint galaxies, thus being likely in a very advanced evolutionary stage, whereas XMMXCS J2215.9-1738 is less massive and shows a deficit of faint galaxies on the red sequence, thus being in a less evolved dynamical state.
At lower redshift ($z \sim 0.3$), \citet{braglia09} studied the star formation activity in two clusters with opposite dynamical states: even if low star formation activity is found in both cluster cores, a star formation activity is found in the less evolved cluster, out to its virial radius and beyond, while no star formation activity is found in the more evolved cluster, thus hinting to a link between the dynamical state and the star formation activity in clusters.
The different dynamical status may explain the conflicting evidence observed at high redshift.

	The evolution of star formation activity within clusters has been studied in many papers \citep[e.g.][and references therein]{haines09}, starting with the pioneering work of \citet{butcher84}.
The latter authors looked at the fraction of blue galaxies, $f_{blue}$, in clusters and its evolution with redshift, thus probing the impact of dense environments on star formation activity.
This seminal study found an increase of $f_{blue}$ in clusters with increasing redshifts -- the so-called Butcher-Oemler effect -- out to $z \sim 0.5$, thus pointing to an accelerated evolution in clusters.
However, it has been subsequently shown that Butcher-Oemler effect studies may be severely affected by methodological biases.
\citet{andreon99} have shown that a strong bias in the \citet{butcher84} cluster sample may account for the seen effect.
According to subsequent works \citep{de-propris04,goto05}, $f_{blue}$ seems not to depend on the cluster mass.
\citet{de-propris03} have shown the necessity to use mass-selected galaxy sample: in fact, if one use an optical luminosity-selected galaxy sample as \citet{butcher84} did, higher redshift samples will be biased towards low mass starburst galaxies, not included in lower redshift samples, leading to an artificial increase of $f_{blue}$.
At last, \citet{andreon06} have shown that the criterium used to define a blue galaxy needs to take into account the younger mean age of the Universe and the secular increase in the star formation rate with redshift.

	Recent studies find no evidence for a Butcher-Oemler effect out to $z \sim 0.5$ \citep{andreon06,haines09}; nevertheless, \citet{andreon08a} found evidence for a Butcher-Oemler effect when comparing a $z \sim 1$ cluster with local clusters.

	We here take advantage of CFHTLS/WIRDS public deep images covering the $z \sim 2.2$ JKCS\,041 cluster \citep{andreon09} to address the aforementioned issues.
JKCS\,041 presents a well defined red sequence \citep{andreon11} populated by a homogenous population of galaxies with extremely synchronized stellar ages \citep{andreon11a} and an extended X-ray emission with $T = 7.3^{+6.7}_{-2.6}$ keV \citep{andreon09,andreon11b}, attesting the presence of a formed potential well, deep enough to be hot and retain the intracluster medium.
JKCS\,041 thus offers a unique opportunity to probe star formation activity in clusters and the Butcher-Oemler effect out to $z \sim 2.2$.

	The plan of this paper is as follows: we describe in Section \ref{sec:data_analysis} the data used for JKCS\,041, along with the analysis led on them.
We estimate in Section \ref{sec:jkcs041_zphot} JKCS\,041 photometric redshift.
We then study the Butcher-Oemler effect in Section \ref{sec:boe} and JKCS\,041 star formation activity in Section \ref{sec:sfa}.
We summarize and discuss our results in Section \ref{sec:discus_concl}.

	In this paper, we adopt $H_0 = 70$ km s$^{-1}$ Mpc$^{-1}$,  $\Omega_m = 0.30$ and $\Omega_\Lambda = 0.70$.
All magnitudes are in the AB system and masses are computed with a \citet{chabrier03} Initial Mass Function (IMF).
JKCS\,041 virial radius, estimated from the X-ray temperature, is $r_{200} = 1.53 \arcmin$  \bibpunct[]{(}{)}{;}{a}{}{;} \citep[][;$\sim 0.76$ Mpc at $z \sim 2.2$]{andreon09} \bibpunct{(}{)}{;}{a}{}{,} and the cluster center is defined as the barycenter of the X-ray emission.


\section{Data and analysis \label{sec:data_analysis}}

\subsection{Data}

	JKCS\,041 is in the $\sim$ 0.6 deg$^2$ area covered by CFHTLS deep survey ($u^*g'r'i'z'$ bands) and by WIRDS follow-up in the infrared filters ($JHK_s$ bands, 50\% point source completeness: $K_s = 24.7$) (Bielby et al., in preparation, catalogs are available on the CFHT Science Data Archive site \footnote{\url{http://www1.cadc-ccda.hia-iha.nrc-cnrc.gc.ca/cfht/WIRDST0002.html}}).
Throughout this work, we use the T0002 release of catalogs generated using $K_s$-band as detection image and the other bands  in analysis mode.
More specifically, we use magnitudes measured in 2\arcsec~apertures (2\arcsec \ are $\sim$ 17 kpc at $z \sim 2.2$) for colours, and ``total'' magnitudes, both corrected for Galactic extinction using \citet{schlegel98}.

	In order to use these catalogs for the study of JKCS\,041, we need: a) to identify stars; b) to correct the underestimate of photometric errors listed in the original catalog; c) to correct for (minor) residual photometric offset;   d) to  measure photometric redshifts and e) to correct for their systematic biases. We detail them in turn.

	The VIMOS VLT Deep Survey project \citep[VVDS,][]{le-fevre05} gives spectroscopic redshifts, $z_{spec}$, of several thousand of objects in the same area.
For our spectroscopic sample, we use objects in common with T0002 catalogs -- rejecting the edgings -- with $K_s \le 23$ (see hereafter) and a secured $z_{spec}$ (flag=3,4), thus yielding a spectroscopic sample of 2537 galaxies ($z_{spec} \lesssim 1.5$) and 366 stars ($z_{spec} = 0$).

\subsection{Star removal}

\begin{figure}
\resizebox{\hsize}{!}{\includegraphics{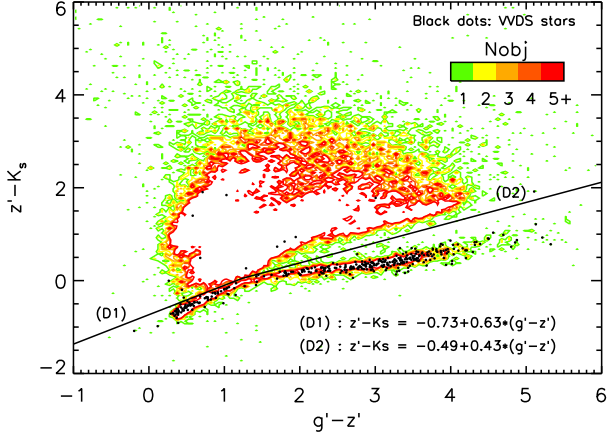}}
\caption{
Star/galaxy separation:  $g'-z'$ vs $z'-K_s$ diagram.
Noisy contours give the density of objects at each location.
Spectroscopically confirmed stars are marked with black dots.
The broken line marks the star/galaxy separation threshold.}
\label{fig:stargal_sep}
\end{figure}

	Stars are identified, and removed, in the colour-colour plane, as in \citet{cowie94} and later works.
In the $z'-K_s$ vs $g'-z'$ colour-colour diagram, spectroscopic-identified stars populate a narrow locus offset from galaxies, as shown in Figure \ref{fig:stargal_sep}. 
We therefore classify as star every object bluer in $z'-K_s$ than the broken line shown in Figure \ref{fig:stargal_sep}.
This criterium excludes more than 95\% of the spectroscopically confirmed stars.
We checked that the latter constitute an unbiased (for our purposes) sample of stars, because they cover the same colour-colour locii of the large and representative sample in \citet{finlator00}.
Stars not identified as such in this phase will be removed in later phases   (during the photometric redshift selection and the  background statistical subtraction phase).

\subsection{Photometric errors correction}

	Because of a slight noise correlation introduced by image resampling during the stacking, flux errors as derived by SExtractor \citep{bertin96}  are underestimated \citep[e.g.][]{casertano00,andreon01}.
Following \citet{andreon01}, by binning the images we can recover the actual background noise, and thus compute the underestimation factor, thanks to the fact that  the correlation is present on small spatial scales only.
We found a factor of 1.5  for optical bands, in agreement with \citet{ilbert06a} and \citet{coupon09}, and a factor 2.0 for near-infrared bands.
The larger factor for  near-infrared bands is due to the native larger pixel size of the images and thus more finely resampling  in the stacking.
After correction of the error underestimate,  completeness, defined as $S/N = 5$, occurs at $K_s=23$ mag, $J = 23.6$ mag and $z'-J= 29.5-1.25 \times K_s$ mag.
Lower $S/N$ data are never used in this work and even more restrictive cuts are used in most instances, as detailed below.

\subsection{Photometric redshift estimation}

	For photometric redshift estimation, we use \textsc{Eazy} \citep{brammer08}  with default settings and a $K_s$-band magnitude prior.
We use both the full photometric redshift probability distribution function, $p(z)$, and $z_{mp}$, the latter as a point estimate of the photometric redshift. $z_{mp}$ is the redshift posterior mean \citep[see $\S$2.5 of ][]{brammer08}.
As noted in previous works \citep[e.g.,][]{brodwin06,ilbert06a,coupon09,ilbert09,barro11}, offsets in the photometric calibration or the inadequacy of the templates to reproduce all the observed SEDs can lead to systematic offsets in the photometric redshift estimation.
\textsc{Eazy} addresses the latter by using a template error function.
In order to fix photometric offsets, we use the subsample of our VVDS galaxies having a $S/N$ greater than 10 in all the 8 photometric bands (2281 galaxies) and we compute the mean difference between the predicted (best fit model) and observed magnitude.
More precisely, we  compute this average one filter at a time, we apply the photometric offsets, and iterate until the procedure converged.
Five iterations were sufficient for convergence and the found offsets are listed in Table \ref{tab:mag_offsets}.
These small shifts, comparable with previous works \citep{ilbert06a,ilbert09,coupon09,barro11}, were applied for photometric redshift estimation only.

\begin{table}
\centering
\caption{Systematic offsets $m_{fit}-m_{meas}$ between measured and best-fit model magnitudes \label{tab:mag_offsets}}
\begin{tabular}{c c c c c c c c}
\hline
\hline
$u^*$	&	$g'$	&	$r'$	&	$i'$	&	$z'$	&	$J$	&	$H$	&	$K_s$	\\
\hline
-0.08		&	0.06	&	0.01	&	0.03	&	-0.01	&	-0.05&	-0.02&	-0.01		\\
\hline
\end{tabular}
\end{table}

\begin{figure}
\resizebox{\hsize}{!}{\includegraphics{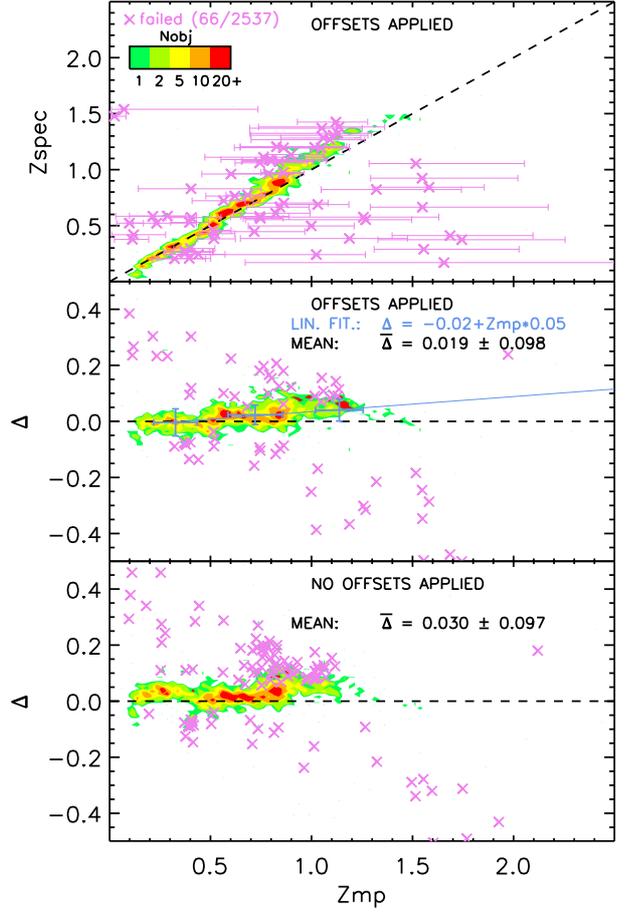}}
\caption{
Performance of the photometric redshift estimate: contours give the density of objects at each location.
In this Figure only, we consider that the photometric redshift estimate fails when $z_{spec}$ does not belong to the 3$\sigma$ confidence interval of $z_{mp}$ (light magenta).
\textit{Upper panel}: $z_{spec}$ vs $z_{mp}$ when offsets are applied.
\textit{Middle panel}: residuals $\Delta = (z_{spec} - z_{mp}) / (1 +z_{mp})$ vs $z_{mp}$ when offsets are applied.
The light blue line is a linear fit to the three points in light blue, representing the median values when the data are binned in three bins.
We observe that $z_{spec} \gtrsim 1$ galaxies tend to have their $z_{mp}$ value underestimated (cf. light blue line).
\textit{Lower panel}: residuals $\Delta$ vs $z_{mp}$ when no offsets are applied.
}
\label{fig:eazy_perf}
\end{figure}

	Figure \ref{fig:eazy_perf} shows $z_{mp}$ vs $z_{spec}$ (top panel) after correction of the photometric offsets.
The scatter is $0.019 \pm 0.098$, much better than if the correction is not applied ($0.030 \pm 0.097$),  largely because an improvement at low redshift, as shown from comparison of residuals before and after correction (see middle and bottom panels).

	Even after correction of the photometric offsets, the photometric redshift tends to underestimate the spectroscopic redshift at $z>1$ (see middle panel),  as already noted by \citet{brammer08}.
The linear fit to the binned data (light blue) illustrates this trend.
We checked that a similar underestimate of the redshift holds for the sub-sample of  red galaxies, defined as $(U-V)^{model}_{rest-frame}>1.3$ mag.
We apply this last correction in the only place of this work where it is needed, in Section \ref{sec:jkcs041_zphot}.

\subsection{Background removal}

	When estimating the properties of JKCS\,041 galaxy population -- in Sections \ref{sec:boe} and \ref{sec:sfa}, we need to account for galaxies on the cluster line of sight, that we generally call background, by a two step procedure: first, we will make a photometric redshift selection by removing galaxies (and QSO) that are at $z<1.7$ or $z>3.5$ at $\ge 99$ \% confidence.
This selection is accomplished by keeping objects with:
\begin{equation}
\int_0^{1.7} p(z)dz \le 0.99 \quad \textnormal{and} \quad \int_{3.5}^{+ \infty} p(z)dz \le 0.99
\label{eq:zphot_crit}
\end{equation}
We note that the latter equation removes only very few objects.
We remark that this selection is very effective in removing "low redshift" galaxies, as it removes 2495 of the 2522 galaxies with $z_{spec} \le 1.5$ in our VVDS sample.
As later detailed, we explored other possible choices, and results are insensitive to the precise used recipe.

As a second step of background subtraction, we use a large control area ($\sim 0.1$ deg$^2$ around the cluster, excluding a disk of 7\arcmin~radius centered on the cluster) to estimate the residual background.
This step also subtracts any star not identified as such by colours.

	We here analyze the impact on our sample selection of wrong photometric redshift estimate.
In Figure \ref{fig:eazy_perf}, a sizable  number of points scatters off from the diagonal in the top panel. 
In this Figure only, we consider that the photometric redshift estimate fails when $z_{spec}$ does not belong to the 3$\sigma$ confidence interval of $z_{mp}$ (galaxies in light magenta).
A large majority of the outliers are objects with fairly large errors. 
Most of the outliers are in the lower-right corner of the $z_{spec}$ vs $z_{mp}$ plot (i.e. galaxies with overestimated photometric redshift).
These galaxies increase the noisiness of our measurements but do not introduce any bias, because kept in the sample. 
The most troublesome ones  are high-redshift galaxies with a largely underestimated redshift (i.e. the objects in the very top-left corner) {\it and} a nominal small redshift uncertainty.
In fact, these galaxies, if they exist, would be a source of incompleteness in our sample, which discards all galaxies which are at $z<1.7$ at 99\% confidence.
In our spectroscopic sample, this very situation never happens; there are only two galaxies -- lying at a redshift ($z_{spec} \sim 1.6$) lower than the one we are interested in -- which are in a qualitative similar situation.


\section{JKCS$\,$041 photometric redshift estimate \label{sec:jkcs041_zphot}}

\begin{figure*}
\centering
\includegraphics[width=0.9\linewidth]{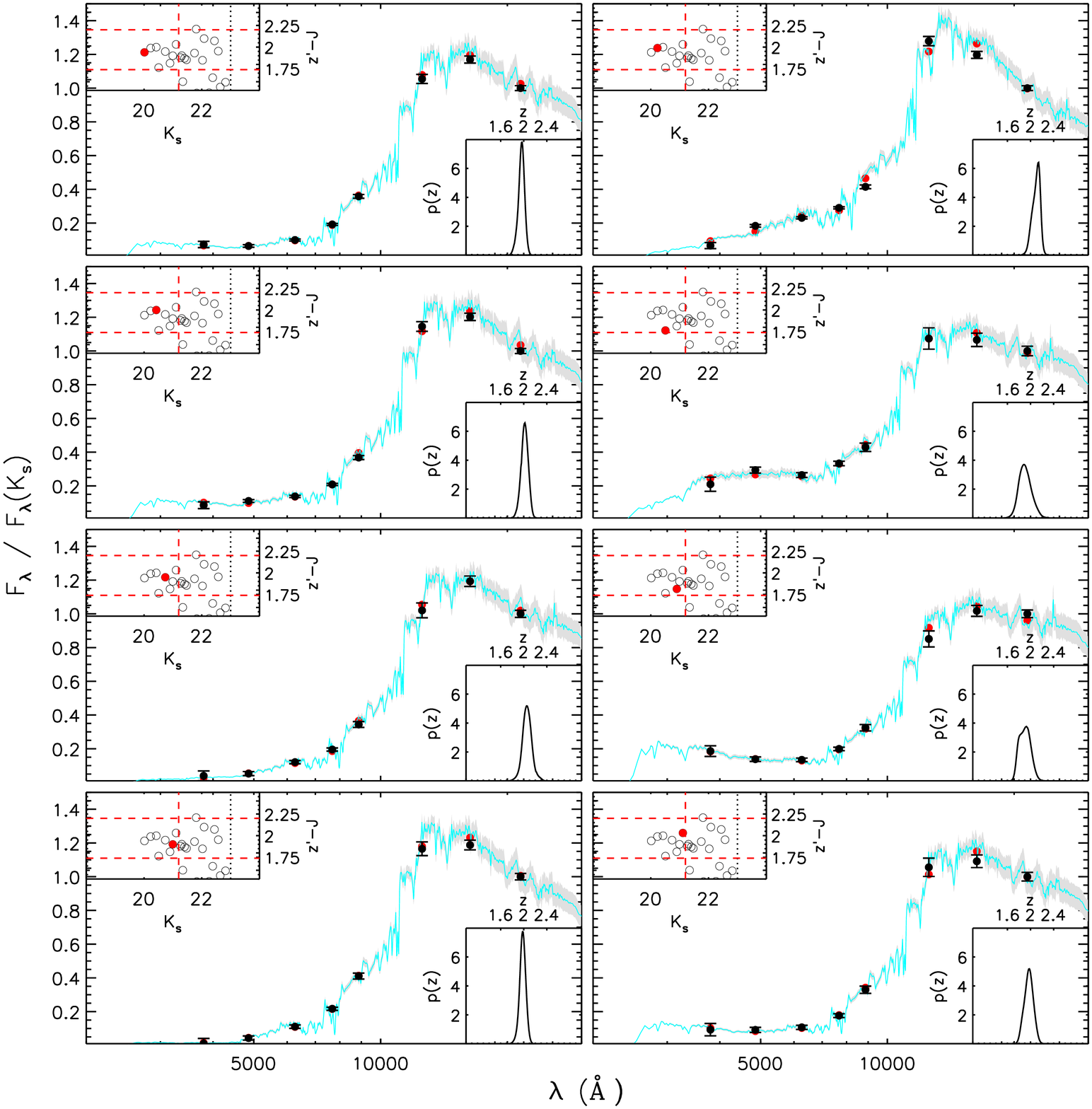}
\caption{
JKCS$\,$041 bright red sequence galaxies analysis.
Large panels shows the normalized SED (black dots with error bars), along with the \textsc{Eazy} best-fit template (cyan line) and the best-fit model photometric points (red dots).
The thin light gray shaded area around the best-fit template shows the model uncertainty in  the best-fit template, as given in \textsc{Eazy}.
\textit{Upper-left insets}: Colour-magnitude relation of all galaxies within $0.5 \times r_{200}$.
The closed (red) point emphasizes the object whose SED is shown.
Red dashed lines represent cuts in magnitude and colour used in this plot, whereas the black vertical dotted line represents the $K_s = 23$ limit.
\textit{Lower-right insets}: photometric redshift probability distribution function $p(z)$ for the considered galaxy.
All those red and bright galaxies have a SED with a prominent 4000 \AA~break  and are well fitted with a template at $z\sim 2.0$.}
\label{fig:pdzsed}
\end{figure*}

	In order to estimate JKCS$\,$041 redshift, we select bright ($K_s \le 21.2$) objects within $0.5 \times r_{200}$ and within 3$\sigma$ from the $z-J$ vs $K_s$  colour-magnitude relation \citep[$1.74 \le z-J \le 2.2$,][]{andreon11a}, because this choice maximizes the cluster membership likelihood (such bright and red galaxies are rare in the field, as measured all around the cluster).
We emphasize that we do not use any photometric redshift selection in this Section.
For each of these 8 galaxies,  Figure \ref{fig:pdzsed} shows their SED  along with the  \textsc{Eazy} best-fit template and, as insets, its position in the colour-magnitude diagram (\textit{upper-left inset}) and  the photometric redshift probability  distribution function $p(z)$ (\textit{lower-right inset}). 
We observe that the fits are of good quality and that those 8 galaxies show a prominent 4000 \AA~break near the $J$ band, characteristic of high-redshift old galaxies.
We remark that the two bluest galaxies have a $p(z)$ less peaked and slightly shifted towards lower redshifts.

\begin{figure}
\resizebox{\hsize}{!}{\includegraphics{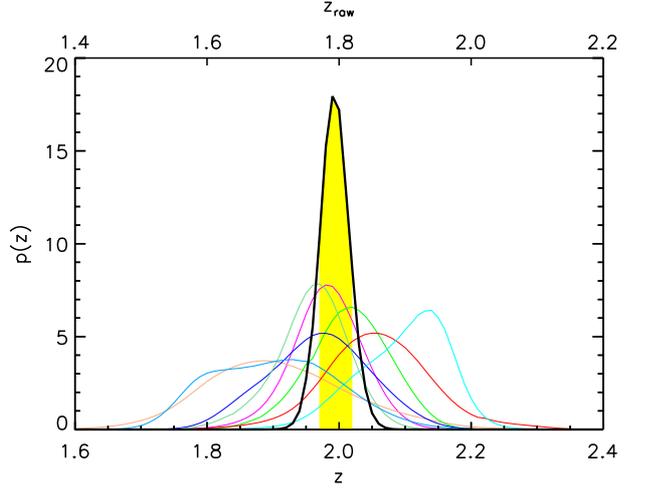}}
\caption{
JKCS$\,$041 photometric redshift estimate : the thick black line shows the cluster photometric redshift, the thin coloured lines represent photometric redshifts of the 8 bright galaxies on the cluster red  sequence shown in Figure \ref{fig:pdzsed}.
The area shaded in yellow represents the 1$\sigma$ confidence interval. 
The top axis shows the raw photometric redshift, not accounting for the known underestimate of photometric redshifts.}
\label{fig:pdz}
\end{figure}

	In Figure \ref{fig:pdz}, we gather the photometric redshift probability distribution functions $p(z)$ for the 8 selected galaxies (thin coloured lines).
All the 8 $p(z)$ peak between $z \sim 1.8$ and $z \sim 2.2$.
If we assume that all those 8 objects belong to the cluster, the cluster photometric redshift is obtained by multiplying the $p(z)$ functions, as they have been derived from independent data.
The result is plotted in Figure \ref{fig:pdz} as a thick black line: it has a very peaked shape around $z \sim 2.0$ (1$\sigma$ confidence interval: $[1.97,2.02]$) in broad agreement with the photometric redshift of $z_{phot} = 2.20 \pm 0.11$, robustly estimated in \citet{andreon11}.
We stress that our photometric redshift estimate is derived using a photometric redshift correction extrapolated from $z \lesssim 1.5$: it is hence slightly less robust than \citet{andreon11} estimate, which uses the red sequence and is calibrated on the cluster ClG J0218.3-0510, which has a spectroscopic redshift $z_{spec} = 1.62$ \citep{papovich10,tanaka10}.

	The top axis of Figure \ref{fig:pdz} shows that JKCS\,041 would have a high redshift ($z_{raw} \sim1.8$)  even when neglecting the correction for the underestimate of photometric redshift shown in Figure \ref{fig:eazy_perf}.
This underestimate hints to a calibration problem for the models. 
Nevertheless, as model predictions match JKCS\,041 colours for a redshift $z \sim 1.8$, we will use in the rest of this article this value to redshift our models.
This approach allows to use predicted colours in agreement with our observations (in particular, see Figure \ref{fig:cmdsfr} where predicted colours for a Simple Stellar Population match the red sequence).


\section{Butcher-Oemler effect \label{sec:boe}}

	The measurement of the Butcher-Oemler effect requires particular attention to the way galaxies are selected in mass and classified as blue/red (see Section \ref{sec:intro}).
First, we use a mass-selected sample.
We select all galaxies more massive than $1.34 \times 10^{11} M_\odot$, which corresponds to our completeness ($S/N = 5$) limit in the worst case (old red galaxies).
By mass we refer, as in previous works, to the  \cite{bruzual03} models (2007 version, CB07 hereafter) model mass, and specifically the mass of the gas that will eventually turn into stars  (see e.g. \citet{longhetti09} for other possible definitions of model mass).
These (model) masses are computed for solar metallicity, formation redshift $z_f=5$, and either Simple Stellar Populations or exponentially declining star-forming $\tau$ model with $0 < $ SFH $\tau$ (Gyr) $\le 10$.

\subsection{Blue/red definition}
	We define a galaxy as blue if it is bluer than a CB07 model with $\tau=3.7$ Gyr, as in \citet{andreon04,andreon06,andreon08a} and \citet{loh08}.
This galaxy will be bluer by 0.2 mag in $B-V$ than red sequence galaxies at the present epoch \citep[which would be a blue galaxy by the original definition of][]{butcher84}. 
The rationale behind this choice is described in detail in \citet{andreon06} and \citet{andreon06a}, but in short we take into account the stellar evolution of galaxies with time.

	We emphasize that a galaxy to be classified as blue should be very blue, at least $0.8$ mag bluer in $J-K_s$ than the red sequence.
This is imposed by the requirement of following galaxies back in time, with a criterium independent of the redshift (under the assumption of exponentially declining SFH).
A significantly narrower selection is precluded because, at lower redshift, it would end up on the top of the red sequence.

\begin{figure} 
\resizebox{\hsize}{!}{\includegraphics{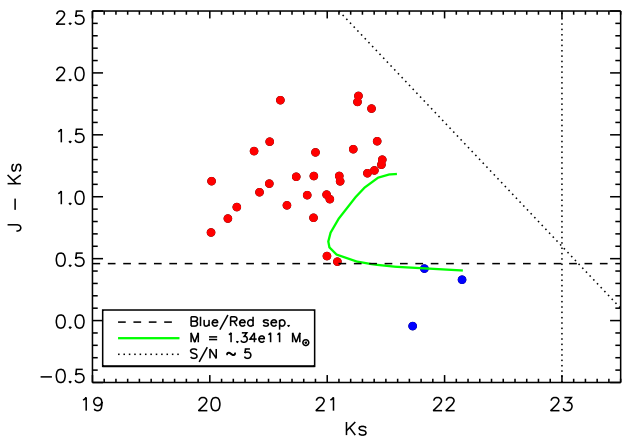}}
\caption{
Colour-magnitude diagram for galaxies in JKCS\,041 line of sight (i.e. JKCS\,041 \textit{and} residual background galaxies) within $2 \times r_{200}$.
The black dotted lines represent the $S/N \sim 5$ completeness.
The green thick solid line represents the mass cut ($M = 1.34 \times 10^{11} M_\odot$).
The dashed line indicates the blue/red threshold.}
\label{fig:cmd}
\end{figure}

	Figure \ref{fig:cmd} shows the $J-K_s$ vs $K_s$ (about rest-frame $u^*-r'$ vs.  $r'$) colour-magnitude diagram for galaxies within $2 \times r_{200}$ of the cluster center and more massive than $1.34 \times 10^{11} M_\odot$.
We adopted this colour index because it best matches the colour index used in our comparison sample below.
As later detailed, our results do not depend on the used colour index.
Not many galaxies are blue (below the dashed line in Figure \ref{fig:cmd}), even before accounting for residual (after photometric redshift selection) background in the JKCS\,041 line of sight.
In addition, we underline that all (resp. more than 90\% of) our sample galaxies have a $S/N \ge 10$ in $K_s$ (resp. $J-K_s$).

\subsection{Measurement of a Butcher-Oemler effect}
	We compute the blue fraction and the number of red and blue members galaxies accounting for residual background galaxies (i.e. in the line of sight, and not belonging to the cluster) using our control field  following the Bayesian methods introduced in \citet{andreon06}.
We adopt uniform priors for the parameters.
We consider three regions, defined by $r/r_{200} \le 0.5$ , $0.5 < r/r_{200} \le 1$ and $1 < r/r_{200} \le 2$.

\begin{figure}
\resizebox{\hsize}{!}{\includegraphics{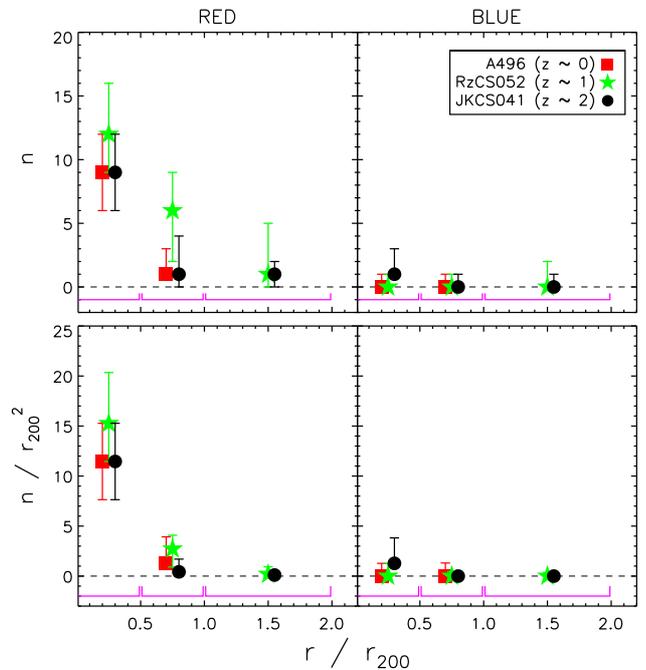}}
\caption{
Radial number profile (upper panels) and radial number density profile (lower panels) of red (left panels) and blue (right panels) member galaxies as a function of the cluster-centric distance, coded as indicated in the legend, after accounting for the background.
The three radial ranges used are indicated by the magenta lines.
Points indicate the maximum a posteriori, the error bars the shortest 68\% interval.
Points are slightly shifted horizontally for readability.}
\label{fig:profile}
\end{figure}

	In order to provide lower redshift reference clusters, we consider RzCS\,052 at $z = 1.016$ \citep{andreon08a}, and Abell 496 \citep[A496 hereafter,][]{abell58} at $z = 0.033$ \citep{struble99}, already used in the Butcher-Oemler study in \citet{andreon08a}, considering a lower mass threshold.
Table \ref{tab:cluster_prop} lists some key characteristics of our cluster sample used to study the Butcher-Oemler effect.
As one can see, our three clusters have roughly similar masses; moreover our study shows that they also have similar richness (see hereafter, Figure \ref{fig:profile}).
We use as colour index $u^*-r$ for A496 and $I-z'$ for RzCS\,052, thus adopting in both cases a rest-frame $u^*$-like band as blue band, as for JKCS\,041.
We recompute the blue fraction of A496 and RzCS\,052 clusters for galaxies with $M \ge 1.34 \times 10^{11} M_\odot$.

\begin{table}
\centering
\caption{Cluster sample used to estimate the Butcher-Oemler effect \label{tab:cluster_prop}}
\begin{tabular}{l |  l l l}
\hline
\hline
Cluster					&	$r_{200}$	&	$\sigma_v$			&	$M_{200}$			\\
						&	(Mpc)	&	(km s$^{-1}$)			&	($10^{14} M_\odot$)		\\
\hline
A496 \tablefootmark{a}		&	1.85		&	$721^{+35}_{-30}$		&	7.5					\\
RzCS\,052 \tablefootmark{b}	&	1.04		&	$710^{+150}_{-150}$	&	4.0					\\
JKCS\,041 \tablefootmark{c}	&	0.76		&	-					&	$4.0^{+5.3}_{-3.3}$		\\
\hline
\end{tabular}
\\
\tablefoottext{a}{\citet{rines05}}
\tablefoottext{b}{\citet{andreon08}}
\tablefoottext{c}{\citet{andreon09,andreon11b}}
\end{table}

	Figure \ref{fig:profile} shows red (left panels) and blue (right panels) radial number profiles (upper panels) and radial density profiles (lower panels) for all three clusters, after accounting for the background.
The three clusters have similar blue and red radial profiles, showing that the three clusters have similar richnesses.
The right panel shows that all these three clusters have a negligible number of blue galaxies within $2 \times r_{200}$.
Unfortunately, the low redshift of A496 prevents us to probe its outer regions.

\begin{figure}
\resizebox{\hsize}{!}{\includegraphics[width=\linewidth]{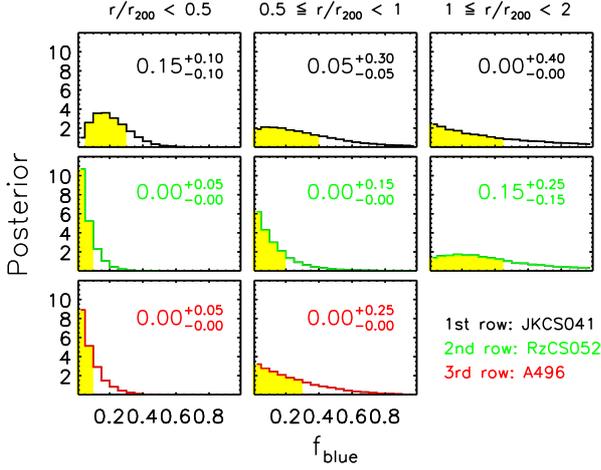}}
\caption{
Posterior probability distribution for the blue fraction for the three clusters (JKCS$\,$041: \textit{upper row}, RzCS$\,$052: \textit{middle row}, A496: \textit{lower row}), in three
radial ranges.
The 68\% shortest confidence intervals are shaded in yellow.}
\label{fig:fblue_hist}
\end{figure}

	Figure \ref{fig:fblue_hist} shows the (posterior) probability distribution of the blue fraction of the three clusters  in the three radial regions (two for A496) and list their summaries (point estimates and 68 \% shortest interval). 
We emphasize that our $f_{blue}$ computation requires the use of the full probability distributions for intervening quantities, $n_{blue}$ and $n_{red}$, not just their point estimates and 68\% uncertainties reported in Figure \ref{fig:profile}.

\begin{figure}
\resizebox{\hsize}{!}{\includegraphics[width=\linewidth]{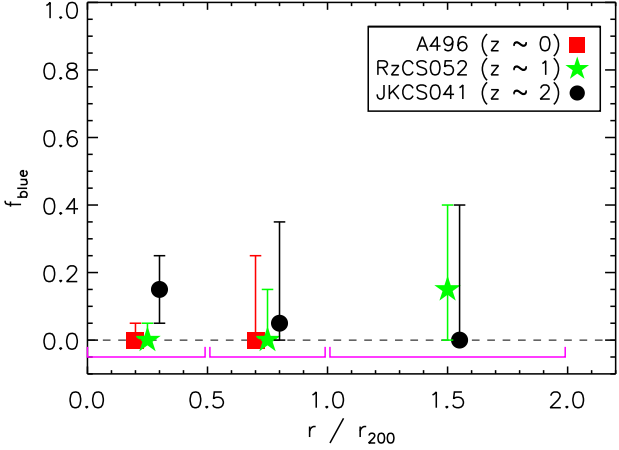}}
\caption{
Blue fraction as a function of the cluster-centric distance: JKCS$\,$041 ($z \sim 2$, black dots), RzCS$\,$052 ($z \sim 1$, green stars), and A496 ($z \sim 0$, red squares).  
Points indicate the maximum a posteriori, the error bars the shortest 68\% interval.
The three radial ranges used are indicated by the magenta lines.
Abscissae are slightly shifted for readability.}
\label{fig:fblue}
\end{figure}

	Figure \ref{fig:fblue} shows the blue fraction profiles of the three clusters and our main result of this section: we observe the same (negligible) amount of blue galaxies more massive than $1.34 \times 10^{11} M_\odot$ all the way up to $z \sim 2.2$ in all radial bins.
The values of $f_{blue}$ are all less than 1.4$\sigma$ away for all the three clusters in all the radial bins.

	To quantify the sensitivity of our data to an evolving $f_{blue}$, we focus on the innermost bin ($r/r_{200} \le 0.5$) of the two clusters at the redshift ends, A496 and JKCS\,041, to maximize the redshift leverage.
By looking at the posterior of $\Delta f_{blue} = f_{blue}^{JKCS041} - f_{blue}^{A496}$, $\Delta f_{blue} \le 0.36$ (resp.  $\Delta f_{blue} \le 0.18$) with 95\% (resp. 68\%) probability.
Hence the rate at which the blue fraction changes, $\Delta  f_{blue} / \Delta z$, is less than 0.16 (resp. 0.08) per unit redshift at 95\% (resp. 68\%) probability.
By a way of comparison, \citet{butcher84} found a slope of 0.5 for a roughly similar radial aperture, but for a sample including less massive galaxies.
Although error bars at $z \sim 2.2$ are large, the redshift leverage of this work allows us to tightly constrain the evolution of $f_{blue}$.

\subsection{Robustness of the result on assumption}
We test the robustness of our result on assumptions:
a) we change the redshift of JKCS\,041 by $\pm0.1$;
b) in place of the selection given in Equation \ref{eq:zphot_crit}, we keep in the sample galaxies with at least 30 \% probability \citep[following][]{tran10} in the $1.5<z<2.1$ range, i.e.
\begin{equation}
\int_{1.5}^{2.1} p(z)dz \ge 0.3
\label{eq:zphot_crit2}
\end{equation}

c) we use the $z'-J$ (rest-frame $\sim$2500-3000 \AA~ - $u^*$)  instead of $J-K_s$ colour index to identify red and blue galaxies.
For those three cases, none of our measurements (fractions and radial profiles) change, not even by 1$\sigma$.
If instead of a mass-selected sample, we use a luminosity-selected sample brighter than an evolved $M_V = -20.8$ mag for all three clusters, then we derive consistent radial profile for the three clusters, i.e. we continue to see no Butcher-Oemler effect.


\section{Star-formation activity \label{sec:sfa}}

\begin{figure}
\resizebox{\hsize}{!}{\includegraphics[width=\linewidth]{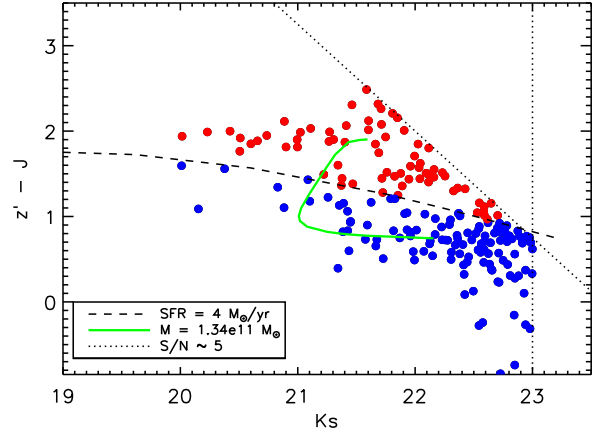}}
\caption{
Colour-magnitude diagram for galaxies in JKCS$\,$041 line of sight (i.e. JKCS\,041 \textit{and} residual background galaxies) within $2 \times r_{200}$.
The black dotted lines represent the $S/N=5$ locii.
The green solid line represents the locus of constant mass $M = 1.34 \times 10^{11} M_\odot$.
The black dashed line indicates the locus of a constant SFR of 4 $M_\odot$ yr$^{-1}$, splitting quiescent and star-forming galaxies.}
\label{fig:cmdsfr}
\end{figure}

	In this section we classify galaxies as star-forming, or quiescent,  from the slope of the UV continuum, specifically according whether they are bluer or redder in the $z'-J$ ($\sim L_{2800}/L_{3700}$) vs $K_s$ diagram than a CB07 model of solar metallicity, $z_{form}=5$ and an exponential declining star formation history with star formation rate (SFR) equals to  4 $M_\odot$ yr$^{-1}$ at $z=1.8$.
This SFR value is chosen according to the analysis of \citet{kriek09}, who studied a $z \sim 2$ spectrum of a quiescent galaxy and found a maximum SFR of 4 $M_\odot$ yr$^{-1}$ (for a Chabrier IMF).
We underline that there is no reference SFR value at $z \sim 2$ used to classify a galaxy as quiescent: we chose the value of \citet{kriek09} study, because it relies on a spectroscopic measurement, though only on one object.
Other works, as \citet{quadri11} for instance, use a different criterium.
In addition to our $M = 1.34 \times 10^{11} M_\odot$ mass-selected sample,  we consider less massive galaxies in a $S/N$ selected sample.
The latter choice has the advantage of enlarging the mass range, but the disadvantage of making difficult, not to say impossible, to compare results derived from data of different depths or for clusters at different redshift.
Figure \ref{fig:cmdsfr} shows the data and the various relevant locii.

\begin{figure}
\resizebox{\hsize}{!}{\includegraphics{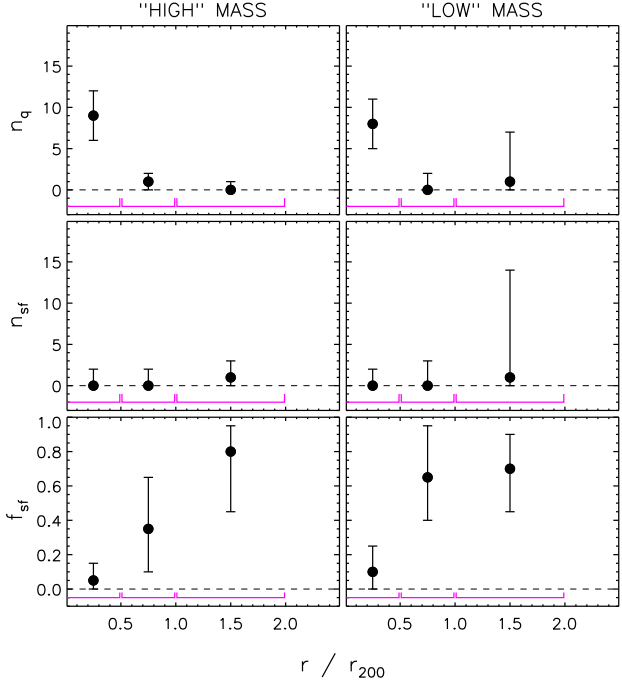}}
\caption{
JKCS\,041 radial profiles of cluster quiescent $n_{q}$ (\textit{upper row}) and star-forming $n_{sf}$ (\textit{middle row}) galaxies, and the cluster star-forming fraction $f_{sf}$ (\textit{lower row}) for the high mass subsample (\textit{left column}) and the less massive subsample (\textit{right column}).
The three radial ranges used are indicated by the magenta lines.
 }
\label{fig:sfr}
\end{figure}

	Using the same Bayesian methods used in previous section for red and blue galaxies, we compute the radial profile of quiescent and star forming galaxies, as well as the radial profile of the star-forming fraction, shown in Figure \ref{fig:sfr}.
This Figure shows a consistent and systematic increase of the fraction of star-forming galaxies with cluster-centric distance for both the $M \ge 1.34 \times 10^{11} M_\odot$ sample (left panels) and the sample of less massive galaxies (right panels).
This is our main result of this section.

\begin{figure}
\resizebox{\hsize}{!}{\includegraphics{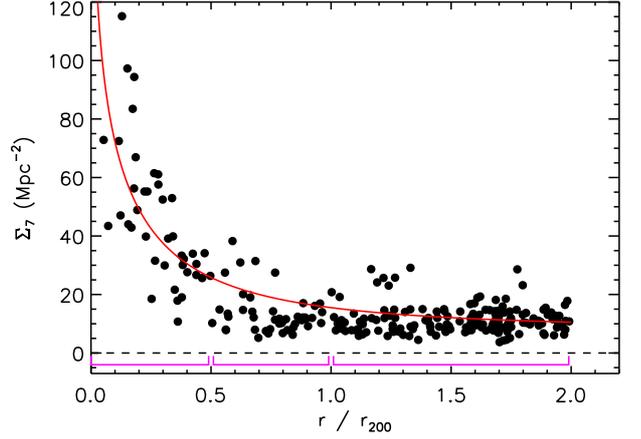}}
\caption{
Local density, estimated from the distance of the 7$^{th}$ nearest neighbor vs JKCS\,041 cluster-centric distance: for this Figure only, galaxies are selected solely with a photometric redshift criterium (Eq. \ref{eq:zphot_crit}) and $K_s \le 23$ mag.
The red line shows the best-fit with a theoretical \citet{navarro96} profile plus a constant to take into account the background.
The three radial ranges used for estimating $f_{blue}$ and $f_{sf}$ are indicated by the magenta lines.
}
\label{fig:density}
\end{figure}

\begin{figure}
\resizebox{\hsize}{!}{\includegraphics{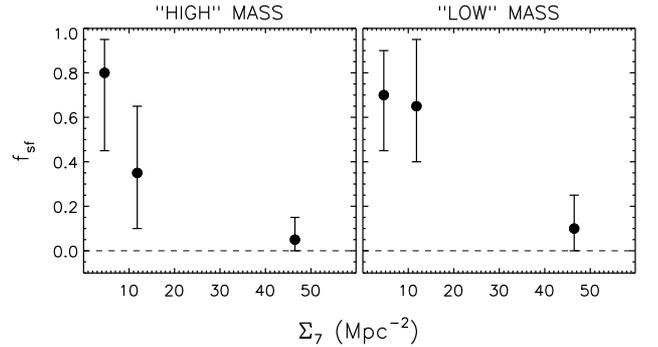}}
\caption{JKCS\,041 star-forming fraction $f_{sf}$ vs density $\Sigma_7$ for the high mass subsample (\textit{left column}) and the less massive subsample (\textit{right column}).
The plotted values for $\Sigma_7$ are the mean values of the best-fit profile in Figure \ref{fig:density} (red line) for each radial range, after background subtraction.
}
\label{fig:sfr_density}
\end{figure}

	Figure \ref{fig:density} shows that local density, estimated from the distance of the 7$^{th}$ nearest neighbour, goes as a function of cluster-centric distance: it  decreases with $r/r_{200}$ until $r/r_{200} \sim 1$, after which its trend cannot be estimated because of uncertainties. 
Note that the cluster center is defined by the X-ray barycenter, not by the peak of the local density itself and thus the peak at low radii is not due to a selection effect.
Since density and cluster-centric distance run hand in hand, at $z \sim 2.2$, the fraction of star-forming galaxies decreases with density.
For illustrative purpose, we plot in Figure \ref{fig:sfr_density} JKCS\,041 star-forming fraction $f_{sf}$ vs background-subtracted density $\Sigma_7$.

	We emphasize that while the high mass sample is complete (in mass), the sample of lower mass galaxies has a completeness depending on whether the galaxy is quiescent or star-forming.
This prevents any quantitative comparison of the star-forming fraction values between the two samples, because, for example, a larger fraction of star-forming galaxies among less massive galaxies may be genuine or just a selection effect,  due to mass-incompleteness of quiescent galaxies of low mass, as already mentioned.
On the other hand, this selection effect is independent on the cluster-centric distance and thus does not affect our conclusion about the increase of the fraction of star-forming galaxies with cluster-centric  distance.

	We test the robustness of our result on assumptions:
a) we change the redshift of JKCS\,041 by $\pm0.1$;
b) we use Eq.\ref{eq:zphot_crit2} for photometric redshift pre-selection;
c) we multiply/divide by 2 the SFR threshold value used to define quiescent/star-forming galaxies;
d) we classify galaxies in star-forming, or quiescent, in the $U-V$ vs $V-J$ plane, as \citet{williams09} and \citet{quadri11}.
We emphasize that in case d) the two classifications are almost identical, but that our rest-frame $J$ photometry comes from extrapolating the available multicolour
photometry.
Our result does not change with case a).
For cases b), c) and d), we observe a radial profile increase with cluster-centric distance, i.e. that the star formation-density relation is already in place at $z \sim 2.2$.
We stress that in all cases, the fraction of massive star-forming galaxies within $r_{200}/2$ is very low ($ \le 15\%$, 1$\sigma$ error bars included).


\section{Discussion and conclusions \label{sec:discus_concl}}

	In this work, we took advantage of CFHTLS ($u^*g'r'i'z'$ bands) and WIRDS ($JHK_s$ bands) images to study the JKCS\,041 cluster.
Using two lower redshift clusters of similar mass (RzCS\,052 at $z = 1.016$ and A496 at $z = 0.032$) as a comparison sample, we studied the evolution with redshift of cluster galaxies properties.
Our results are:
\begin{enumerate}
\item By using photometric redshifts, we confirm that JKCS\,041 is a high redshift cluster, in agreement with \citet{andreon11}.
Our estimate is $z_{phot} = 2.00^{+0.02}_{-0.03}$ (1$\sigma$ errors), after a systematic correction of $\delta z = 0.2$ extrapolated from $z \lesssim 1.5$.
\item Working on a mass-limited sample ($M \ge 1.34 \times 10^{11} M_\odot$) and taking into account the star aging with decreasing redshift, we observe the same (negligible) fraction of blue galaxies all the way up to $z \sim 2.2$ in all radial bins within $2 \times r/r_{200}$ : we thus do not observe any evidence of any Butcher-Oemler effect between $z \sim 2.2$ and $z \sim 0$.
With the forced choice of our definition of blue galaxy, very few galaxies more massive than $M = 1.34 \times 10^{11} M_\odot$ are blue, for all redshifts and radii. 
Although error bars are large, the redshift leverage of this work is at least twice larger than any previous work, allowing us to reject with confidence a change greater than $\Delta  f_{blue} / \Delta z = 0.16$ in the cluster center.
\item JKCS\,041 shows a consistent and systematic increase of the fraction of star-forming galaxies with cluster-centric distance for both the $M \ge 1.34 \times 10^{11} M_\odot$ sample and a sample of less massive galaxies.
In particular, very few (less than 15\%) star-forming galaxies are found within $r_{200}/2$ among high mass galaxies.
\item Density decreases with increasing cluster-centric distance for $r \le r_{200}$, and thus our statements above may be rephrased in terms of local density.
\end{enumerate}

	\citet{andreon08a} led a very similar analysis of the Butcher-Oemler effect for RzCS\,052 and A496, the only difference being that a lower mass cut is used ($M \ge 4 \times 10^{10} M_\odot$).
Interestingly, they find an evolution in the fraction of blue galaxies between $z \sim 1$ and $z \sim 0$, whereas we do not.
Gathering their and our results hints towards a \textit{downsizing}-like scenario \citep[e.g.][]{cowie96,treu05,iovino10,peng10}, where the most massive galaxies have their properties set in the very early Universe ($z \gg 1$), while less massive galaxies still evolve in $0 < z < 1$.

	Our observations of JKCS\,041 show that most of bright core cluster galaxies are red and passive at $z \sim 2.2$, an observation that models still have difficulties to match \citep[e.g.][]{menci08,romeo08}.

	The presence of a star formation-density relation in JKCS\,041 is in agreement with the works of \citet{chuter11} and \citet{quadri11}  at $z \lesssim 1.8$ and those on the XMMU J2235-2557 cluster at $z = 1.39$ \citep{lidman08,rosati09,strazzullo10,bauer11} and in disagreement with studies on the XMMXCS J2215.9-1738 cluster at $z = 1.46$ \citep{hayashi10,hilton10}.

	This variety of results may be either a manifestation of a spread of star formation-density relations at high redshift -- perhaps due to the cluster dynamical status, or just the result of unidentified systematics.
Enlarging the sample used for the Butcher-Oemler effect -- and particularly selecting likely low redshift descendants of JKCS\,041 -- and deepening the analysis of the star formation activity in JKCS\,041 is necessary to strengthen our results, and will be addressed in future works.
However, JKCS041 is uniquely adapted for galaxy evolutionary studies, even if it has a photometric redshift only, because it is so far the only $z \gtrsim 1.5$ cluster with measured intra-cluster medium properties -- hence $r_{200}$ --  \textit{and} a sizable red population.
More clusters, with robust estimate of mass are needed to consolidate the link, put forth by JKCS041, between star formation and environment at high redshift.




\begin{acknowledgements}

We acknowledge financial contribution from the agreement ASI-INAF I/009/10/0 and from Osservatorio Astronomico di Brera.

Based on observations obtained with
MegaPrime/MegaCam\footnote{The full text acknowledgement is at
http://www.cfht.hawaii.edu/Science/CFHLS/cfhtlspublitext.html} and 
WIRCAM\footnote{The full text acknowledgement is  at
http://ftp.cfht.hawaii.edu/Instruments/Imaging/WIRCam/WIRCamAcknowledgment.html}
at CFHT. 

\end{acknowledgements}

\end{document}